\documentstyle[12pt]{article}
\begin{document}

\thispagestyle{empty}
\setcounter{page}0

{}~\vfill

\begin{center}
{\Large\bf A queer reduction of degrees of freedom}

\vfill

{\large L. V. Avdeev\footnote{ E-mail: $avdeevL@thsun1.jinr.dubna.su$
 \protect\\ Supported in part by the RFFR grant \# 94-02-03665-a.}}
and~
{\large M. V. Chizhov\footnote{ Permanent address: Center of Space
Research and Technologies, Faculty of Physics, University of Sofia, 1126
Sofia, Bulgaria. E-mail: $physfac2@bgearn.bitnet$ }}

\vspace{1cm}

{\em Bogoliubov Laboratory of Theoretical Physics, Joint Institute for
Nuclear Research, $141\,980$ Dubna $($Moscow Region$)$, Russian
Federation}

\end{center}

\vfill

\begin{abstract}
The classical dynamics of antisymmetric second-rank tensor matter fields
is analyzed. The conformally invariant action for the tensor field leads
to a positive-definite hamiltonian on the class of the solutions that
are bounded at the time infinity (plane waves). Only the longitudinal
waves contribute to the energy and momentum. The helicity proves to be
equal to zero.

\end{abstract}

\vfill

\newpage

In the field theory an elementary particle with an integer (a half
integer) spin $s$ is usually described by a totally symmetric tensor
field $\Phi_{\mu_1...\mu_{\scriptstyle s}}$ (a tensor spinor field
$\psi_{\mu_1...\mu_{{\scriptstyle s}-1/2}}$) \cite{Schwinger}. The
tensors (spinor-tensors) symmetrized by other Young tableaux were also
considered in the literature. Two forms of the action are known for the
antisymmetric second-rank tensor fields. In the first case the action
possesses the gauge symmetry~\cite{gauge}:

\begin{equation} {\cal S}_A = -\int {\rm d}^4 x~
 \Big[~ \frac 1 4 (\partial_\lambda A_{\mu\nu})~
  \partial^\lambda A^{\mu\nu}
  -\frac 1 2 (\partial_\mu A^{\mu\lambda})~ \partial^\nu A_{\nu\lambda}
 \Big] , ~~~~
 \delta A_{\mu\nu} = \partial_\mu \Lambda_\nu - \partial_\nu \Lambda_\mu
 . \label{A}
\end{equation}
The gauge fields (\ref{A}) appear in supergravity theories in diverse
dimensions \cite{diverse} and in the effective low-energy field theory
derived from relativistic strings~\cite{strings}. These fields have been
well studied, and eventually, a unitary $S$-matrix was constructed even
in the nonabelian case~\cite{Frolov}. The gauge invariance leaves only
one (longitudinal) polarization state for the tensor gauge field, as
differs from the ordinary vector gauge fields.

Another possibility is the conformally invariant action for the
antisymmetric tensor matter fields $T_{\mu\nu}$:
\begin{equation} {\cal S}_T = \int {\rm d}^4 x~
 \Big[~ \frac 1 4 (\partial_\lambda T_{\mu\nu})~
  \partial^\lambda T^{\mu\nu}
  -(\partial_\mu T^{\mu\lambda})~ \partial^\nu T_{\nu\lambda}
 \Big] . \label{T}
\end{equation}
Such an action naturally arises in the conformal field theory
\cite{Mintchev} and in conformal supergravity~\cite{de Wit}. However,
the dynamics of such fields has not been completely investigated. There
is an assertion that, in the Euclidean space, $T_{\mu\nu}$ describe
three physical and three ghost degrees of freedom~\cite{Fradkin}. We are
going to study the classical dynamics of the antisymmetric tensor matter
fields in the Minkowski space. If we limit ourselves to the solutions
that are bounded at the time infinity, then the hamiltonian proves to be
positive-definite; just two degrees of freedom contribute to the energy
and momentum.

Six independent components of the real tensor field $T_{\mu\nu}$ can be
parameterized by a three-dimensional vector $A_i$=$T_{0i}$ and a
pseudovector $B_i$=${1\over 2}\epsilon_{ijk}~ T_{jk}$, where the Latin
indices take on values 1, 2, 3, and are summed over when repeated.
Following the standard formalism for free classical fields \cite{BSh},
let us work in the momentum representation
\begin{equation}
 {\bf A}(x) = \int \frac {{\rm d}^4 k} {(2\pi)^{3/2}}~ \exp(i~kx)~
  {\bf A}(k),~~~~~
 {\bf B}(x) = \int \frac {{\rm d}^4 k} {(2\pi)^{3/2}}~ \exp(i~kx)~
  {\bf B}(k). \label{Fourier}
\end{equation}
Choose a special reference frame ${\bf e}_i$:~ ${\bf e}_i \cdot {\bf
e}_j = \delta_{ij}$,~ $[{\bf e}_i \times {\bf e}_j] = \epsilon_{ijk}~
{\bf e}_k$,~ ${\bf e}_3 = {\bf k}/|{\bf k}|$.~ Let the decomposition of
the fields over the basis be ${\bf A}(k) = a_i(k)~ {\bf e}_i$,~ ${\bf
B}(k) = b_i(k)~ {\bf e}_i$. Then the action, eq.~(\ref{T}), can be
expressed as
\begin{eqnarray}
 {\cal S}_T &=& \pi \int {\rm d}^4 k~ \Big\{ \sum_{i=1}^2
 \left[ a_i^*(k) \big( k_0^2 +{\bf k}^2 \big) a_i(k)
  + b_i^*(k) \big( k_0^2 +{\bf k}^2 \big) b_i(k)
 \right] \nonumber \\
&& +~2 k_0~ |{\bf k}|~
 \big[ a_1^*(k)~ b_2(k) +b_2^*(k)~ a_1(k) -a_2^*(k)~ b_1(k)
  -b_1^*(k)~ a_2(k)
 \big] \nonumber \\
&& +~a_3^*(k) \big( k_0^2 -{\bf k}^2 \big) a_3(k)
 + b_3^*(k) \big( k_0^2 -{\bf k}^2 \big) b_3(k) \Big\} .
 \label{T1}
\end{eqnarray}

An additional rotation
\begin{eqnarray*}
&a_1(k) = \frac 1 {\sqrt 2} \big[ c_1(k) +d_2(k) \big] , ~~~~~
 a_2(k) = \frac 1 {\sqrt 2} \big[ c_2(k) +d_1(k) \big] , ~~~~~
 a_3(k) = c_3(k);& \\
&b_1(k) = \frac 1 {\sqrt 2} \big[ d_1(k) -c_2(k) \big] , ~~~~~
 b_2(k) = \frac 1 {\sqrt 2} \big[ d_2(k) -c_1(k) \big] , ~~~~~
 b_3(k) = d_3(k)~&
\end{eqnarray*}
diagonalizes action (\ref{T1})
\begin{eqnarray}
 {\cal S}_T &=& \pi \int {\rm d}^4 k~ \Big[
 c_1^*(k) \big( k_0 -|{\bf k}| \big) ^2 c_1(k)
 + c_2^*(k) \big( k_0 +|{\bf k}| \big) ^2 c_2(k) \nonumber \\
&& +~c_3^*(k) \big( k_0 -|{\bf k}| \big) \big( k_0 +|{\bf k}| \big)
 c_3(k) ~~+~(c\to d)~ \Big] . \label{T2}
\end{eqnarray}
The principle of extreme action leads to the following field equations
\begin{eqnarray}
& \big( k_0 -|{\bf k}| \big) ^2 c_1(k_0,{\bf k}) = 0, ~~~~~~~
 \big( k_0 +|{\bf k}| \big) ^2 c_2(k_0,{\bf k}) = 0,& \nonumber \\
& \big( k_0 -|{\bf k}| \big) \big( k_0 +|{\bf k}| \big) ~
 c_3(k_0,{\bf k}) = 0.& \label{eq}
\end{eqnarray}
The same equations hold for $d_i(k)$. The general solutions to
eqs.~(\ref{eq}) are of the form
\begin{eqnarray}
 c_1(k_0,{\bf k}) &=& \delta \big( k_0 -|{\bf k}| \big) ~ c_1({\bf k})
 +\delta' \big( k_0 -|{\bf k}| \big) ~ \widetilde{c}_1({\bf k}),
 \nonumber \\
c_2(k_0,{\bf k}) &=& \delta \big( k_0 +|{\bf k}| \big) ~ c_2({\bf k})
 +\delta' \big( k_0 +|{\bf k}| \big) ~ \widetilde{c}_2({\bf k}),
 \nonumber \\
c_3(k_0,{\bf k}) &=&
 \delta \big( k_0 -|{\bf k}| \big) ~ \overline{c}_3({\bf k})
 +\delta \big( k_0 +|{\bf k}| \big) ~ c_3({\bf k}). \label{solution}
\end{eqnarray}

We want to note here that the transverse components $c_1(k_0,{\bf k})$
and $d_1(k_0,{\bf k})$ involve only positive frequencies $k_0$=$|{\bf
k}|$. Other transverse components $c_2(k_0,{\bf k})$ and $d_2(k_0,{\bf
k})$ involve only negative frequencies $k_0$=$-|{\bf k}|$.

The fact that ${\bf A}(x)$ and ${\bf B}(x)$ are real leads to the
relations
\begin{eqnarray}
 c_2({\bf k}) = s~ c_1^*(-{\bf k}) -c~ d_1^*(-{\bf k}), \nonumber \\
\widetilde{c}_2({\bf k}) = -s~ \widetilde{c}_1{}^*(-{\bf k})
 +c~ \widetilde{d}_1{}^*(-{\bf k}), &&
 \overline{c}_3({\bf k}) = -c_3^*(-{\bf k}), \nonumber \\
d_2({\bf k}) = c~ c_1^*(-{\bf k}) +s~ d_1^*(-{\bf k}), \nonumber \\
\widetilde{d}_2({\bf k}) = -c~ \widetilde{c}_1{}^*(-{\bf k})
 -s~ \widetilde{d}_1{}^*(-{\bf k}), &&
 \overline{d}_3({\bf k}) = -d_3^*(-{\bf k}), \label{real}
\end{eqnarray}
where $c = {\bf e}_1({\bf k}) \cdot {\bf e}_1(-{\bf k})
= - {\bf e}_2({\bf k}) \cdot {\bf e}_2(-{\bf k})$,~~
$s = {\bf e}_1({\bf k}) \cdot {\bf e}_2(-{\bf k})
= {\bf e}_2({\bf k}) \cdot {\bf e}_1(-{\bf k})$,~~ $c^2+s^2=1$.~
Thus, $\overline{c}_3$, $\overline{d}_3$, and all the amplitudes with
index 2 are not independent and can be eliminated.

The energy--momentum 4-vector for the tensor matter field is defined as
$$ {\cal P}_\mu = \int {\rm d}^3 {\bf x}~
 \Big[ (\partial_\mu T_{\alpha\beta})
  \frac {\partial {\cal L}_T} {\partial (\partial_0 T_{\alpha\beta})}
  -g_{\mu 0} {\cal L}_T
 \Big] . $$
By a direct calculation we find
\begin{eqnarray}
 {\cal P}_0 &=& \int {\rm d}^3 {\bf x}~ \Big[ ~
 \frac 1 2 (\partial_0 {\bf A})^2 -\frac 1 2 (\partial_i {\bf A})^2
 +(\partial_i A_i)^2 ~~+~({\bf A}\to {\bf B})~ \Big] \nonumber \\
&=& \int {\rm d}^3 {\bf k}~ \Big\{
 \widetilde{c}_1{}^*({\bf k})~ \widetilde{c}_1({\bf k})
 -|{\bf k}|~
 \big[ \widetilde{c}_1{}^*({\bf k})~ c_1({\bf k})
  +c_1^*({\bf k})~ \widetilde{c}_1({\bf k})
 \big] \nonumber \\*
&& +~2~ {\bf k}^2~ c_3^*({\bf k})~ c_3({\bf k})~~ +~(c\to d)~ \Big\} ,
 \label{E}
\end{eqnarray}
\begin{eqnarray}
{\cal P}_i &=& \int {\rm d}^3 {\bf x}~ \Big\{
 (\partial_i {\bf A}) \cdot (\partial_0 {\bf A})
 +(\partial_i {\bf B}) \cdot (\partial_0 {\bf B})
 +2 (\partial_i {\bf A}) \cdot [\vec{\partial} \times {\bf B}] \Big\}
 \nonumber \\
&=& \int {\rm d}^3 {\bf k}~ k_i~ \Big\{
 \widetilde{c}_1{}^*({\bf k})~ c_1({\bf k})
 +c_1^*({\bf k})~ \widetilde{c}_1({\bf k})
 +2~ |{\bf k}|~ c_3^*({\bf k})~ c_3({\bf k})~~ +~(c\to d)~ \Big\} .
 ~~~~~~~ \label{P}
\end{eqnarray}

The obtained energy and momentum cannot be made simultaneously diagonal.
The straightforward diagonalization of eq.~(\ref{E}) reveals two
positive- and two negative-energy transverse modes which cannot be
interpreted as relativistic particles. These facts are due to the
presence of $\delta'$-type solutions in eqs.~(\ref{solution}). In the
co-ordinate representation (\ref{Fourier}) such solutions do not
correspond to ordinary plane waves but rather grow linearly with the
time and are unbounded at the infinity. If we set their amplitudes to
zero, that is, choose to restrict ourselves to the plane-wave sector,
then the transverse plane-wave amplitudes drop out of the
energy--momentum. Everything is reduced to the longitudinal waves. Both
energy (\ref{E}) and momentum (\ref{P}) can be written uniformly
\begin{equation}
 {\cal P}_\mu = \int {\rm d}^3 {\bf k}~ k_\mu~
 \big[ c_3^+({\bf k})~ c_3^-({\bf k}) + d_3^+({\bf k})~ d_3^-({\bf k})
 \big] , \label{P_}
\end{equation}
where, as usual \cite{BSh}, $c_3^+({\bf k})$=$c_3^*({\bf k}) / \sqrt{2
k_0}$, $c_3^-({\bf k})$=$c_3({\bf k}) / \sqrt{2 k_0}$, $k_0$=$|{\bf
k}|$. The hamiltonian ${\cal P}_0$ becomes positive-definite on the
plane-wave solutions. This important property would not take place, if
one introduced a mass term like $T^{\mu\nu} T_{\mu\nu}$ in
eq.~(\ref{T}). Thus, the free tensor field describes massless
relativistic particles.

The helicity --- the projection of the spin onto the direction of motion
--- proves to be equal to zero. In fact, even without the restriction to
plane waves, the 3-vector of spin
\begin{equation}
 {\bf S} = \int {\rm d}^3 {\bf x}
 \Big( [{\bf A} \times \partial_0 {\bf A}]
  + [{\bf B} \times \partial_0 {\bf B}]
  +\big[ {\bf A} \times [\vec{\partial} \times {\bf B}] \big]
  -\big[ {\bf B} \times [\vec{\partial} \times {\bf A}] \big]
 \Big) \label{spin}
\end{equation}
vanishes on solutions (\ref{solution}).

There are global transformations of the vector and pseudovector fields
\begin{equation}
 \pmatrix{ {\bf A'} \cr {\bf B'} } =
 \pmatrix{ \cos \alpha & \sin \alpha \cr
          -\sin \alpha & \cos \alpha }
 \pmatrix{ {\bf A} \cr {\bf B} } ,~~~~~~
 \delta T_{\mu\nu} = {1\over 2}~ \epsilon_{\mu\nu\kappa\lambda}~
 T^{\kappa\lambda}~ \delta \alpha \label{axial}
\end{equation}
that leave action (\ref{T}) invariant. They induce a conserved axial
charge
\begin{eqnarray}
 {\cal Q} &=& \int {\rm d}^3 {\bf x} ~
 \Big( {\bf A} \cdot \partial_0 {\bf B}
  -{\bf B} \cdot \partial_0 {\bf A}
  -{\bf A} \cdot [ \vec{\partial} \times {\bf A} ]
  -{\bf B} \cdot [ \vec{\partial} \times {\bf B} ]
 \Big) \nonumber \\
&=& -~i \int {\rm d}^3 {\bf k} ~
 \big[ c_1^* ~ \widetilde{d}_1 -\widetilde{d}_1{}^* ~ c_1
  +d_1^* ~ \widetilde{c}_1 -\widetilde{c}_1{}^* ~ d_1
  +2~ |{\bf k}|~ (c_3^*~ d_3 - d_3^*~ c_3)
 \big] ,~~~~~ \label{Q}
\end{eqnarray}
which in the plane-wave sector is reduced to
\begin{equation}
 {\cal Q} = -i \int {\rm d}^3 {\bf k} ~
 \big[ c_3^+({\bf k})~ d_3^-({\bf k}) -d_3^+({\bf k})~ c_3^-({\bf k})
 \big] . \label{Q_}
\end{equation}
Again, the transverse components fall out. The axial (chiral) symmetry
(\ref{axial}) --- as well as the positive definiteness of the energy ---
forbids the mass term for $T_{\mu\nu}$.

As the dynamical invariants for bounded solutions do not depend on the
transverse components of the tensor field, it is natural to suppose that
only the longitudinal excitations are physical. In contrast to the
tensor gauge field (which has only one degree of freedom on shell
\cite{gauge}), two physical states (the longitudinal components of the
vector ${\bf A}$ and pseudovector ${\bf B}$) are left for the tensor
matter field. On the mass shell the transverse components play no role.
They become important as interactions with other fields and a
self-interaction are added~\cite{int}.

It is also worth comparing the tensor matter field with the vector gauge
field. The transverse plane-wave components of the tensor field are
similar to the time and longitudinal polarizations of the photon.
However, in the latter case there is a gauge symmetry which is
responsible for the cancellation of unphysical degrees of freedom.
Introducing the scalar (anti)ghost fields with unusual commutation
rules, one can write the more general BRST transformations \cite{BRST}
which leave the action invariant even after gauge fixing. The ghost
degrees of freedom are subtracted from the total number, thus leaving
only physical degrees of freedom.

In the case of the tensor matter field there is no gauge invariance.
However, like for the tensor gauge fields, we can add a pyramid of ghost
fields (ghosts for ghosts): anticommuting $C_\mu$, $\overline{C}_\mu$
with the ghost numbers $\pm$1, commuting $D$, $\overline{D}$ with the
zero ghost number, and $E$, $\overline{E}$ with the ghost numbers
$\pm$2. Then, naively counting the degrees of freedom, $6 - 2 \times 4 +
2 \times 2$, we get just two longitudinal physical degrees of freedom
for the tensor matter field. The action for the ghosts is of the form
\begin{equation}
 {\cal S}_{\rm ghost} = \int {\rm d}^4 x~
 \Big[ (\partial^\mu \overline{C}{}^\nu)~ \partial_\mu C_\nu
  +(\partial^\mu \overline{D})~ \partial_\mu D
  +(\partial^\mu \overline{E})~ \partial_\mu E
 \Big] . \label{ghost}
\end{equation}
The total action ${\cal S}_T + {\cal S}_{\rm ghost}$ is left invariant
under the following nilpotent transformations with the anticommuting
constant parameter $\Lambda$ whose ghost number equals~1:
\begin{eqnarray}
&\delta T_{\mu\nu} = -(\partial_\mu \overline{C}_\nu
  - \partial_\nu \overline{C}_\mu)~ \Lambda, ~~~~~
 \delta C_\nu = (\partial^\mu T_{\mu\nu} +\partial_\nu D)~ \Lambda,&
 \nonumber \\
&\delta \overline{D} = \partial^\nu \overline{C}_\nu~ \Lambda, ~~~~~
 \delta \overline{C}_\nu = \partial_\nu \overline{E}~ \Lambda, ~~~~~
 \delta E = -\partial^\nu C_\nu~ \Lambda,~~~~~
 \delta D = \delta \overline{E} = 0&
 \label{BRST}
\end{eqnarray}
These transformations are valid for the theory of the free fields. In
case of an interaction, eqs.~(\ref{BRST}) should be generalized to
involve the coupling constants.

\vspace{7mm}

\noindent {\bf Acknowledgements} \vspace{3mm} \nopagebreak

We thank I. V. Tyutin for a detailed conversation and helpful remarks,
M. Yu. Kalmykov for discussions and questions. M. Chizhov is grateful to
N. Karchev and R. Rashkov for a guide in the literature and useful
discussions. He also thanks the Bogoliubov Laboratory of Theoretical
Physics for kind hospitality and the Bulgarian Ministry of Education,
Science and Culture for financial support under Grant-in-Aid for
Scientific Research F-214/2096.

\pagebreak[4]

\end{document}